# Entrainment of heterogeneous glycolytic oscillations in single cells

Anna-Karin Gustavsson, Caroline B. Adiels, Bernhard Mehlig, and Mattias Goksör*

*Department of Physics, University of Gothenburg, SE-41296 Gothenburg, Sweden*

*mattias.goksor@physics.gu.se

Cell signaling, gene expression, and metabolism are affected by cell-cell heterogeneity and random changes in the environment. The effects of such fluctuations on cell signaling and gene expression have recently been studied intensively using single-cell experiments. In metabolism heterogeneity may be particularly important because it may affect synchronisation of metabolic oscillations, an important example of cell-cell communication. This synchronisation is notoriously difficult to describe theoretically as the example of glycolytic oscillations shows: neither is the mechanism of glycolytic synchronisation understood nor the role of cell-cell heterogeneity. To pin down the mechanism and to assess its robustness and universality we have experimentally investigated the entrainment of glycolytic oscillations in individual yeast cells by periodic external perturbations. We find that oscillatory cells synchronise through phase shifts and that the mechanism is insensitive to cell heterogeneity (robustness) and similar for different types of external perturbations (universality).

**Introduction**

Cellular signaling, gene expression, and metabolism are determined by chemical reactions within the cell. The discrete nature of molecular reactions as well as environmental fluctuations and heterogeneity cause fluctuations in these processes. The effect of such noise on cell signaling and gene expression has recently been studied intensively using single-cell analysis[1-5]. However, despite its importance the role of noise and heterogeneity in metabolism[6] is not yet well understood. Heterogeneity is very important in systems where cell-cell communication may cause the cells to synchronise their metabolic oscillations. Cell-cell communication is important because it is a prerequisite for organisation of cell communities and is necessary for evolution to proceed from unicellular to multicellular behaviour.

One of the most intensively studied metabolic dynamics is that of glycolytic oscillations in yeast cells *Saccharomyces cerevisiae*[7-10], for a review see[11]. Remarkably, yeast cells in dense populations appear to be able to synchronise their glycolytic oscillations[9, 12-15] and macroscopic oscillations have been observed in yeast populations[7-10]. This effect appears quite universal in that different signaling molecules/receptors can cause synchronisation[9, 12-19]. It is thus expected that similar synchronisation mechanisms are at work in other oscillatory cell types, possibly via different metabolic species. However, the empirical data pertain to dense cell cultures of millions of cells and previous attempts to investigate the oscillatory behaviour in individual cells have proven challenging[8, 20, 21]. Since observations of macroscopic oscillations do not distinguish between oscillation and synchronisation such measurements neither allow to deduce the microscopic mechanism of synchronisation nor to investigate the role of cell-cell variations.

There is a long history of theoretical investigations of oscillations in glycolytic networks[22, 23] and it is now well understood that sustained oscillations may occur under a wide range of conditions[24-29]. Synchronisation, by contrast, appears to be much more difficult to achieve in computer simulations of model systems[24-27]. Whether or not synchronisation occurs appears to very sensitively depend upon the model parameters[24]. This is at variance with the apparent robustness of the phenomenon observed in dense cell cultures.

In order to understand the synchronisation mechanism and to elucidate the role of heterogeneity it is necessary to quantify how individual cells respond to periodic perturbations of their environments, either due to an externally applied perturbation or through interactions with other cells. To achieve a qualitative and quantitative understanding of the phenomenon requires us to answer three fundamental questions.

First, what is the mechanism of synchronisation? Experimental studies of macroscopic oscillations indicate that phase synchronisation may play a role[15]: Figure 3(A) in[15] shows that the macroscopic phase shift due to a sudden change in environmental conditions depends upon the value of the macroscopic phase just before the perturbation, see also [9]. In order to quantify the effect and to unequivocally establish whether the synchronisation can be achieved by phase changes alone it is necessary to follow how an individual cell is entrained by a periodic perturbation. One must determine whether and at which frequencies cells may continue to oscillate after the perturbation has been switched off, and, most importantly, whether the amplitude changes during entrainment. Last but not least an important question is how the frequency and amplitude of oscillation in the absence of perturbations affect the propensity of the cell to be entrained when the periodic perturbation is switched on. Existing experiments[9, 14, 15] do not answer these questions because (i) only the macroscopic response is measured, and (ii) the response is only measured at a single perturbation pulse. Some model calculations show that mainly non-oscillatory cells become entrained by perturbations[30]. A very important open question is how the phase of an entrained cell relates to the phase of the perturbation. Do cells typically oscillate in phase with the perturbation or not? Theoretical models can show in-phase or out-of-phase entrainment, sensitively depending on model parameters[26]. Macroscopic experiments do not allow resolving this question, because subpopulations oscillating out-of-phase will only lead to a lowering of the amplitude of the macroscopic signal. In order to determine the mechanism of synchronisation, these experiments must be performed on individual cells.

Second, how robust is the synchronisation mechanism? In a theoretical model for phase synchronisation, the efficiency of the mechanism is determined by the heterogeneity of the cells as well as the strength of the entrainment[31]. This is very important because no two cells are alike, and different cells respond differently to external perturbations. In order to quantify how this heterogeneity may prevent synchronisation, we must study the response of an ensemble of independent individual cells with different properties. Can a periodic perturbation entrain cells that initially oscillate at substantially different frequencies? Theoretical analysis using a recent model has shown that perturbations with the synchronising metabolite acetaldehyde (ACA) are not sufficient to entrain cells when they have different frequencies[32]. Most importantly, in an ensemble of many cells, how large fraction of the individual cells becomes entrained? To answer this question single-cell studies are required, because it is impossible to distinguish between full and partial synchronisation in experiments that only measure macroscopic properties. In order to determine how quickly the mechanism entrains it is necessary to investigate responses of individual cells over many subsequent periods of the periodic perturbation. The degree of synchronisation is quantified by computing an order parameter[31].

Third, how universal is the synchronisation mechanism? Entrainment involves the entire glycolytic network, and not just a single reaction or chemical species. The effect might in fact be the result of the combined response to several different chemical species[32]. The kinetics leading to synchronisation is thus very complicated, but entrainment appears to occur for a wide range of different conditions and types of perturbations[9, 12-19]. The question is whether the mechanism leading to synchronisation is the same for all species.

In order to answer the questions raised above we report on the results of novel experiments subjecting individual yeast cells to periodic perturbations. Using an experimental setup that combines an optical trap for cell positioning with microfluidics for precise spatial and temporal control of the extracellular environment we induce glycolytic oscillations in individual cells and investigate their response to periodic perturbations. These experiments allow us to address the questions raised above.

To enforce strict periodicity of the perturbations, a microfluidic device is used to achieve complete and reversible changes of the extracellular milieu. This makes it possible to experimentally investigate how the phase, the frequency and the amplitude of an individual-cell oscillation are affected by periodic external perturbations. Analysing these parameters from individual cells allows us to determine the mechanism of entrainment, and thus to answer the first question raised above.

To answer the second question, we calculate the order parameter of the phases of the oscillations of the individual cells. This reveals the degree of synchronisation. Studying individual cells allows us to determine which fraction of cells becomes entrained by the perturbation. To ensure that no additional perturbations from cell-cell interactions complicate the analysis, we position the cells far apart from each other by means of an optical trap. Cell-cell interactions are further reduced by flushing out intercellular signaling agents by the flow in the microfluidic chamber[28].

To answer the third question we investigate through which biochemical reactions in the glycolytic reaction network the perturbation causes entrainment. Starting out, we investigated the individual cell responses to periodic addition of ACA to gain information about the single-cell responses at conditions previously studied in populations[13, 15] (see Supplementary Fig. S1 online). Strikingly, we found that removal of cyanide without any addition of ACA caused a similar response, in contrast to the results shown in previous population studies[9]. In this study we therefore chose to investigate the effect of periodic cyanide perturbations. Induction of oscillations has mostly been achieved by addition of glucose and subsequent addition of 2-8 mM cyanide[26, 33, 34]. The most common explanation for macroscopic oscillations to be detected only for this cyanide concentration range is twofold. Firstly, in this range respiration is inhibited by cyanide binding to cytochrome c oxidase. Secondly, in this range the ACA concentration is lowered to levels where the cells become sensitive to ACA secretion by other cells[9, 13, 15, 33-35]. The reason is that cyanide binds ACA and forms lactonitrile[36].

## Results

**Entrainment by cyanide removal.** Alternating between a solution containing 20 mM glucose + 5 mM cyanide and a solution containing only glucose, we show that periodic removal of cyanide entrains the oscillations of individual cells in a similar way as addition of ACA (Fig. 1a and Supplementary Fig. S1 online). In control experiments where only the flow rates in the microfluidic flow chamber are changed but the concentration of chemicals in the solutions are kept constant, no entrainment can be seen (Fig. 2a). The results of the control experiment are in agreement with a previous study, which showed that the oscillatory behaviour is affected only by changes of chemicals and not by changes of flow rates in the microfluidic flow chamber[37].

**Mechanism of synchronisation.** To investigate the mechanism responsible for the observed entrainment, the instantaneous phases were calculated (Figs. 1b and 2b) by means of a standard procedure based upon evaluating the Hilbert transform of the signal[38]. To reveal temporal information about the synchronisation mechanism, the phases before and 9 s after cyanide removal were plotted for perturbations 1, 2, 4 and 8 (Fig. 3a-d). Data from experiments where entrainment is induced by cyanide removal is shown in red, while data from control experiments is shown in black. The corrected phase shifts at the first perturbation were fitted with second-degree polynomials (Fig. 4 and Table 1). The cells in the control experiments show the expected, constant phase shift, independent of the phase of the oscillations at the perturbation. The cells exposed to the periodic cyanide removal, on the other hand, show a significant phase dependence of the phase shift. The results show that cells that have phases close to $-\pi$ or $\pi$ become negatively phase shifted, while cells that have phases close to that of the perturbation remain relatively unaffected by the perturbation. Three important conclusions can thus be drawn from these graphs; (1) there is a phase dependence on the phase shifts in the experiment where cells become entrained (2) most of the entraining phase shifts occur during the first perturbations and at the consequent perturbations the cells have already become entrained (3) there is no phase dependence on the phase shift in the control experiment.

Phase synchronisation assumes that the physical properties of the entrained oscillator remain essentially unchanged and that entrainment is due to phase shifts caused by the perturbation. To test whether the frequencies of the oscillations become affected by the perturbations, the mean and standard deviation of the frequencies were calculated before and after the perturbation interval (Fig. 6a). No significant difference could be seen between the experiments where entrainment was induced and the control experiment, neither with regard to the actual frequencies of the oscillations, nor with regard to the distribution of the frequencies. The amplitudes of the oscillations also remained relatively unaffected by the perturbation for oscillatory cells. However, also non-oscillatory cells became entrained by the perturbation (Fig. 5). For this response to be detectable, also the amplitude must become affected. This indicates that phase changes are not sufficient to entrain non-oscillatory cells. In contrast to results of model calculations[26], under no circumstances could entrainment out-of-phase with the periodic perturbation be observed, neither for oscillatory nor non-oscillatory cells.

**Robustness of the synchronisation mechanism.** The degree of synchronisation is quantified by the so-called order parameter $r(t)$ (see Methods). It increases very rapidly in the perturbation interval and already within the first minute of perturbations, the order parameter reaches a value close to unity, corresponding to complete synchrony (Fig. 1c). In the absence of external perturbation the order parameter is expected to tend to zero, assuming that there are no cell-cell interactions that could stabilise synchronisation. However, this is true only provided that the number of cells $N$ used in the analysis is large. In general, we expect the order parameter to be of the order of $1/\sqrt{N}$, corresponding well with the measured value in the absence of perturbations (Figs. 1c and 2c).

How quickly the order parameter decays after the periodic perturbation is switched off depends upon the distribution of the frequencies of the individual cells. We expect that the order parameter decays as $r(t) = r_0 e^{-\Delta \omega t}$, see Methods. This expectation is borne out by our experiments (Fig. 6b-d). Note also that since the oscillations are initially induced simultaneously in all cells, it is expected that the order parameter should be high at the start of the experiment and then decay if there are no cell-cell interactions (Figs. 1c and 2c). The decay of the order parameter at the start of the experiment cannot, however, be well described simply by the distribution of frequencies at the start of the experiment. It can instead be explained by a measured difference in duration of the initial spike of NADH, causing cells with the same frequency to become out of phase already within the first oscillation period.

**Universality of the synchronisation mechanism.** It remains to determine how cyanide can work as a synchronising agent and whether this mechanism is likely to be universal. There are several possible explanations as to why removal of cyanide may have a synchronising effect. One reason could be that respiration is no longer inhibited when cyanide is removed. To cause the observed response, which occurs within a few seconds, it would be necessary that cyanide be released very rapidly from its interaction with cytochrome c oxidase. To investigate whether respiratory reactions are responsible for the synchronisation, the experiment was repeated by changing from glucose/cyanide to glucose using a hypoxic glucose solution. However, also here entrainment could be seen (data not shown). This indicates that the main cause of entrainment at cyanide removal is not due to respiratory reactions. However, care has to be taken when performing experiments using hypoxic media, since yeast cells can respire also in very oxygen poor environments. Even low concentrations of oxygen in the hypoxic medium could cause cell responses when cyanide is removed.

A second reason for the response could be that the pH is altered when cyanide is added. In previous studies, Poulsen *et al.* induced oscillations using 4 mM KOH and hypoxic solutions instead of cyanide[35]. In their studies, the addition of KOH caused a similar pH increase as the addition of cyanide, raising the pH value 0.06 units. Measurements of the pH of our solutions showed that addition of 5 mM cyanide increased the pH value with 0.09 units. To determine whether the cells respond to such a change in pH, the measurement was repeated with perturbations with a hypoxic glucose solution without cyanide, where the pH was adjusted using KOH. This experiment showed that the cells still became entrained by a change from glucose/cyanide to glucose at hypoxic conditions, also when the pH was kept constant (data not shown). It thus seems unlikely that the pH difference is causing the response at cyanide removal.

A third reason for the observed synchronisation could be cyanide binding to intracellular ACA. The phase dependence of the phase shift differs for different chemical species[13]. Our data reveals that the phase response to cyanide removal is comparable to previous studies of synchronisation by ACA addition in populations[15] and in single cells (Supplementary Fig. S1 online), while population studies of perturbations with e.g. glucose reveals a different phase relation[9]. This supports the hypothesis that the removal of cyanide causes a similar response as addition of ACA.

## Discussion

The experiments reported here show that the mechanism behind the synchronisation of individual oscillatory yeast cells is phase synchronization rather than frequency or amplitude modulation. Our data show that the strongest phase response is achieved when the concentration of NADH is at a minimum, where the removal of cyanide prolongs the subsequent period roughly by one third. Oscillatory cells continue to oscillate with frequencies and frequency distributions largely unaffected by the periodic perturbation. Also the amplitude of the oscillations remains relatively unaffected by the perturbations. These results support the paradigm originally put forward by Kuramoto and his collaborators (for a review see[39]): that periodic perturbations can entrain oscillators by affecting just the phases of the oscillators, resulting in a robust and universal mechanism. Theoretical models tend to be very sensitive to changes in parameter values and using different parameters can result in in- or out-of-phase synchronisation of the oscillations. In contrast to some model results[26], our results show no cases of entrainment out-of-phase, neither for oscillatory nor non-oscillatory cells.

Our experiments show that non-oscillatory cells also become entrained by external periodic perturbations, so these cells must also contribute to macroscopic oscillations observed in e.g. Ref. 40. However, in contrast to the entrainment of oscillatory cells, which relies solely on phase synchronisation, entrainment of non-oscillatory cells requires a change of the amplitude (as well as the frequency). In this case the entrainment cannot be explained solely in terms of phase shifts.

The graphs in Figure 3 reveal temporal information about the entrainment of the individual cells, where the clustering of phases can be seen to occur already during the first few perturbations. This is consistent with the rapid increase of the order parameter (Fig. 1c). Heterogeneity is present in the phase response of individual cells, but despite the heterogeneity the majority of oscillatory cells eventually become entrained by the perturbation. The final phase distribution is within 20% of the original distribution. This shows that the synchronisation mechanism is very robust with regard to cell heterogeneity. This result is in stark contrast with results of model calculations[24, 25, 30].

Our results indicate that cyanide can work as a synchronising agent due to its ability to bind intracellular ACA. Removal of cyanide will then have the same effect as addition of ACA, where the response is transduced to the first part of glycolysis via NADH/NAD$^+$ and ATP/ADP[9, 40, 41]. Here it is thus impossible to completely separate the cell responses due to cyanide removal and ACA addition. Different externally applied chemical species thus appear to synchronise individual cells by similar mechanisms, giving further support to the hypothesis that the synchronisation mechanism is universal. Intracellular cyanide reactions are rather complex. That cyanide can act in more ways than by binding ACA and inhibit respiration has been indicated in previous studies, where the oscillatory tendency was stronger when cyanide was added than under hypoxic conditions[34] and under hypoxic conditions where ACA was flushed away by a flow[42]. And indeed, recently Hald *et al.* showed that cyanide also reacts with other metabolites, namely pyruvate and dihydroxyacetone phosphate (DHAP), and that cyanide might affect the behavior of glycolytic oscillations in more ways than just by binding ACA and inhibiting respiration[36]. Previous studies have also shown that cyanide causes longer trains of oscillations than other inhibitors of respiration, such as antimycin A and azide[43, 44] and that oscillations disappear if both cyanide and azide are present[35]. The role of cyanide inhibiting respiration by binding to cytochrome c oxidase and the contribution of respiratory reactions to the oscillatory behavior have recently been discussed by Schrøder *et al.*, who found that oscillations disappear for strains with deletions of a gene coding for subunit VI of cytochrome c oxidase[45]. This finding was unexpected, since the general assumption is that cytochrome c oxidase is completely inhibited at cyanide concentrations used for oscillation studies and should not contribute at all to the oscillatory behavior of the cells. Cyanide reactions with glucose has, however, been shown to be negligible in the pH range of our experiments[36]. These results highlight the need for further studies to fully understand intracellular cyanide reactions.

In summary, we report on experimental results that allow us to determine the mechanism, the robustness, and the universality of synchronisation of glycolytic oscillations in yeast cells. Our results pose challenges to theoretical modeling, namely to reproduce our results by simulations of the dynamics of glycolytic reaction networks, and to explain the robustness and universality of the synchronisation mechanism. A successful model must show (1) the phase responses to periodic perturbations reported in this work, (2) the robustness of phase synchronisation also for the cell heterogeneity reported here, and (3) the universality of the mechanism; a wide range of periodic perturbations entrain the oscillations. Our experiments are the first that show entrainment of individual yeast cells by periodic perturbations. They support the paradigm originally put forward by Kuramoto: that periodic perturbations can entrain oscillators by affecting just the phases of the oscillators, resulting in a robust and universal mechanism.

## Methods

**Cell preparation.** Yeast cells (*Saccharomyces cerevisiae*, X2180), were grown in a medium comprising 10 g/l glucose, 6.7 g/l yeast nitrogen base and 100 mM potassium phthalate (pH 5.0) on a rotary shaker at 30°C until glucose was exhausted, as described in[46]. The cells were harvested by centrifugation (3500 g for 5 min), washed twice in 100 mM potassium phosphate buffer (pH 6.8) and subsequently glucose starved in the buffer solution on a rotary shaker for 3 h at 30°C. The cells were then washed in the buffer solution and stored at 4°C until use.

**Experimental procedure.** The experimental setup and procedure was described in detail in[42]. The cells were introduced into a microfluidic flow chamber with four inlet channels[42] and positioned at the bottom of the chamber in arrays of 5x5 cells with a cell-cell distance of 10 μm using optical tweezers[42, 47, 48]. By positioning the cells far apart, cell-cell interactions by e.g. ACA was avoided[28]. The flows in the channels were laminar and solutions from different channels could only mix by diffusion. The extracellular environment in the cell array area could thus be controlled both spatially and temporally by adjusting the flow rates in the different inlet channels, causing the cell array area to be covered by the chemicals from the intended channel. A complete change of environment was achieved within 4 s of a change in flow rates. All channels contained 100 mM potassium phosphate buffer (pH 6.8) and the cells were introduced in the lowest channel. Glycolytic oscillations were induced by a 20 mM glucose + 5 mM KCN solution, introduced in the second highest channel for 10 min before the periodic perturbations started and for 15 min after the perturbations had ended. Entrainment was investigated by periodically introducing the solution of interest in the top channel for 20 s and the glucose/cyanide solution for 20 s, giving a total period time of the perturbation of 40 s. The perturbations continued for 15 periods (10 min). For perturbations with ACA, a solution comprising 20 mM glucose + 5 mM KCN + 1 mM ACA was used. For experiments performed with hypoxic solutions, the solutions were bubbled through with $N_2$-gas for 10 min before the experiment. The cell responses were studied by imaging the NADH autofluorescence from the individual cells using an epi-fluorescence microscope (DMI6000B, Leica Microsystems) and an EM-CCD camera (C9100-12, Hamamatsu Photonics), where images were acquired every other second[42].

**Data analysis.** The acquired images were processed and analysed as described previously[42]. The resulting signals of the NADH fluorescence intensity of individual cells were then analysed using MATLAB (The MathWorks, Inc.). Running averages of the NADH signals were calculated using a window of approximately two periods (55 data points) and were subtracted from the NADH signals to reduce spurious trends and drift (Supplementary Fig. S2 online). This facilitated extraction of the phases of the oscillations. The instantaneous phases of the oscillations of the individual cells, $\phi_n(t)$, were extracted from the NADH signals by means of a standard procedure that is based upon evaluating the Hilbert transform of the signals in MATLAB[38]. The phases are defined in the interval [$-\pi$, $\pi$] and a phase of 0 represents a maximum in the concentration of NADH (Figs. 1b and 2b). In Figures 1 and 2, the ten cells that showed the most stable oscillations before the perturbation where chosen. The phases before, $\phi_{before}$, and 9 s after cyanide removal, $\phi_{after}$, were plotted for perturbations 1, 2, 4 and 8, for the experiments where entrainment was induced (Fig. 3a-d, $N$=20, red dots) and for control experiments where only the flow rates were changed but not the chemicals in the solutions (Fig. 3a-d, $N$=32, black dots). The 9 s delay was chosen to allow the cells time to respond to the perturbation.

The oscillation frequency of the individual cells, $\omega_n$, was calculated in time intervals 0-5, 5-10 min and 20-25 min as the frequency at the maximum peak of the Fourier transforms of the NADH signals multiplied by $2\pi$, and is given as mean ± standard deviation for experiments where entrainment was induced (Fig. 6a, $N$=24, red bars) and for the control experiments (Fig. 6a, $N$=20, black bars).

The phase shifts of the data sets at the first perturbation were calculated as $\Delta\phi = \phi_{after} - \phi_{before}$. If $\Delta\phi > \pi$, $\Delta\phi$ was adjusted with $\pm 2\pi$ to move it into the interval [$-\pi$, $\pi$]. The phase shifts were then corrected by subtracting the expected phase shifts of each cell, $\omega_n \Delta t$, where $\Delta t$ is the time between the measurement of the phase before and after the perturbation. This corrected phase shift was fitted by a second-degree polynomial on the form:

$$\Delta\phi - \omega_n \Delta t = c_2 \phi^2 + c_1 \phi + c_0, \qquad (1)$$

where the parameter values are estimated as mean values and their uncertainty (due to the spread of data points and the finite sample size) is expressed in terms of their 95% confidence interval (Fig. 4 and Table 1).

The degree of synchronisation was characterised by the order parameter $r(t)$, defined as[31, 49]

$$r(t) = \left| \sum_{n=1}^{N} e^{i\phi_n(t)} \right|, \qquad (2)$$

where $N$ is the total number of cells in the experiment ($N$=10 in Figs. 1, 2 and 6c-d, and $N$=14 in Fig. 6b). An order parameter close to unity indicates a high degree of synchronisation, while an order parameter close to zero indicates large heterogeneity in phases among the individual cells and thus low entrainment by the external, periodic perturbation.

When the cells are independent and there is no external perturbation, the order parameter is expected to decay as

$$r(t) = r_0 e^{-\Delta\omega t}, \qquad (3)$$

where $r_0$ is the order at $t = 0$ and $\Delta\omega$ is the standard deviation of $\omega_n$. The frequencies were calculated in time interval 20-25 min and the decay was set to start 18 s after the end of the last perturbation (Fig. 6b-d).

## Acknowledgements
We acknowledge the financial support from the Swedish Research Council to M.G. and to B.M., the Seventh Framework Programme UNICELLSYS to M.G., Stiftelsen Längmanska kulturfonden to A.-K.G, and from the Göran Gustafsson Foundation for Research in Natural Science and Medicine to B.M..

## Author contributions
A.-K.G planned and performed the experiments, analysed the data and wrote the manuscript. All authors discussed the results and commented on the manuscript. C.B.A, B.M. and M.G provided guidance throughout.

## Additional information
**Supplementary information** accompanies this paper at http://www.nature.com/scientificreports

**Competing financial interests:** The authors declare no competing financial interests.

**How to cite this article:** Gustavsson, A.-K., Adiels, C.B., Mehlig, B. & Goksör, M. Entrainment of heterogeneous glycolytic oscillations in single cells. Sci. Rep. 5, 9404; DOI:10.1038/srep09404 (2015).

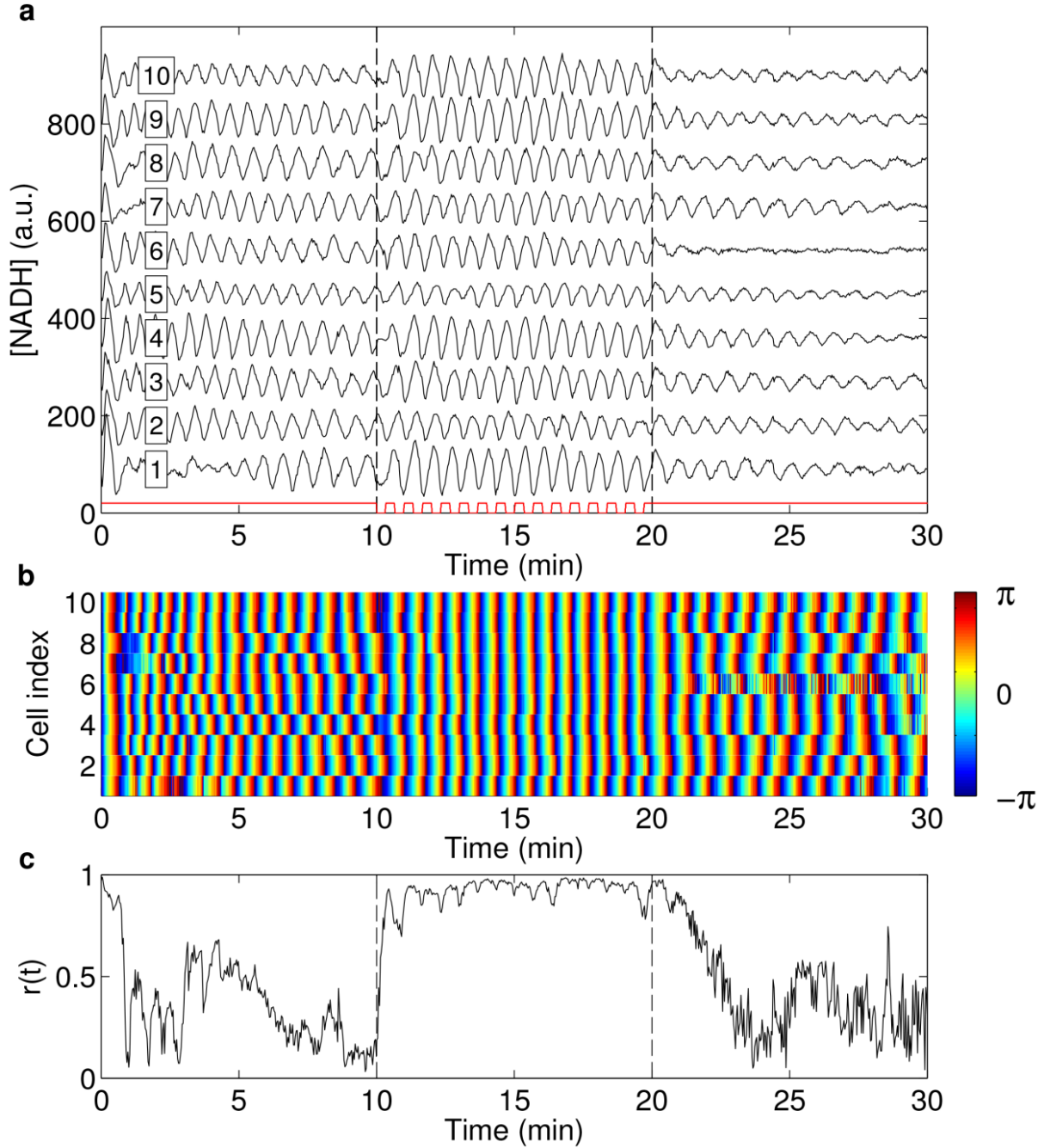

**Figure 1 Entrainment of oscillations by removal of cyanide.** (a) Graph showing the NADH fluorescence intensity from individual cells (cell index in box), where each trace is offset 90 units for clarity. Oscillations are induced at time 0 of the experiment by exposing the cells to a solution containing 20 mM glucose + 5 mM cyanide. After 10 min, cell entrainment by removal of cyanide is investigated by periodically changing the flow rates in the inlet channels of the microfluidic flow chamber, causing a periodic change between the glucose/cyanide solution and a solution containing only 20 mM glucose with a period time of 40 s for a total of 15 periods. The time intervals during which the cells are exposed to cyanide are marked in red. As can be seen, the amplitude of the oscillations remains relatively unaffected by the perturbation. (b) Graph showing the phases of the oscillations of the individual cells shown in (a). Here it can be seen that the phases are shifted as the oscillations become entrained by the perturbation. (c) Graph showing the time dependent order parameter, calculated from the phases shown in (b). An order parameter close to unity indicates synchronisation, while a low value indicates a large heterogeneity of the phases. Here it is thus clear that the cells become entrained by the periodic removal of cyanide through phase shifts.

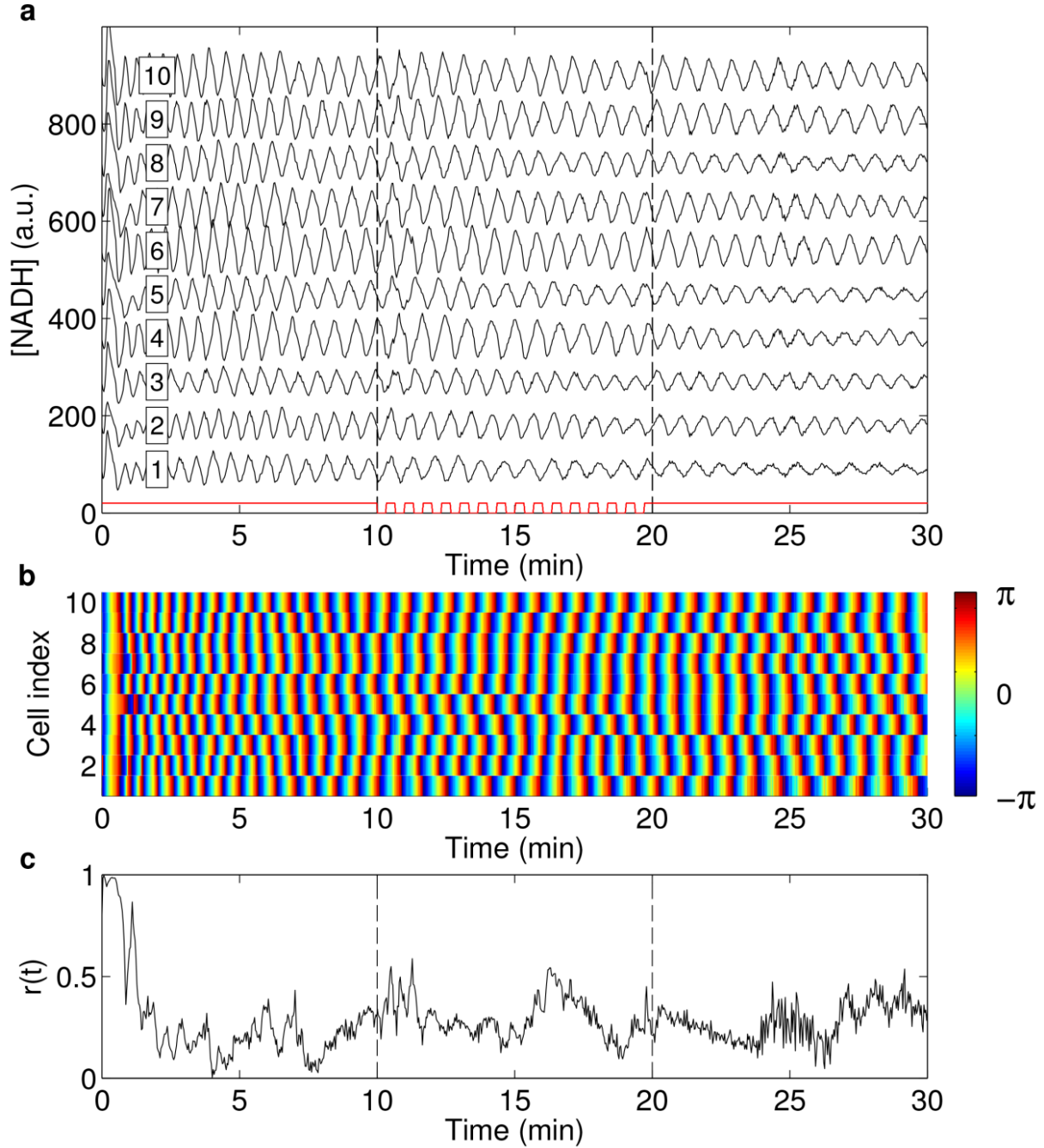

**Figure 2 No entrainment of oscillations by flow rate changes.** (**a**) Graph showing the NADH fluorescence intensity from individual cells (cell index in box), where each trace is offset 90 units for clarity. Oscillations are induced at time 0 of the experiment by exposing the cells to a solution containing 20 mM glucose + 5 mM cyanide. After 10 min, cell entrainment is investigated by periodically changing the flow rates in the inlet channels of the microfluidic flow chamber with a period time of 40 s for a total of 15 periods, while keeping the concentration of the chemicals in the solutions constant. The time intervals during which the flow rates are changed are marked in red. As can be seen, the amplitude of the oscillations remains unaffected by the perturbation. (**b**) Graph showing the phases of the oscillations of the individual cells shown in (**a**). Here no shift of the phases can be seen when the flow rates are changed. (**c**) Graph showing the time dependent order parameter, calculated from the phases shown in (**b**). An order parameter close to unity indicates synchronisation, while a low value, as shown here, indicates large heterogeneity of the phases.

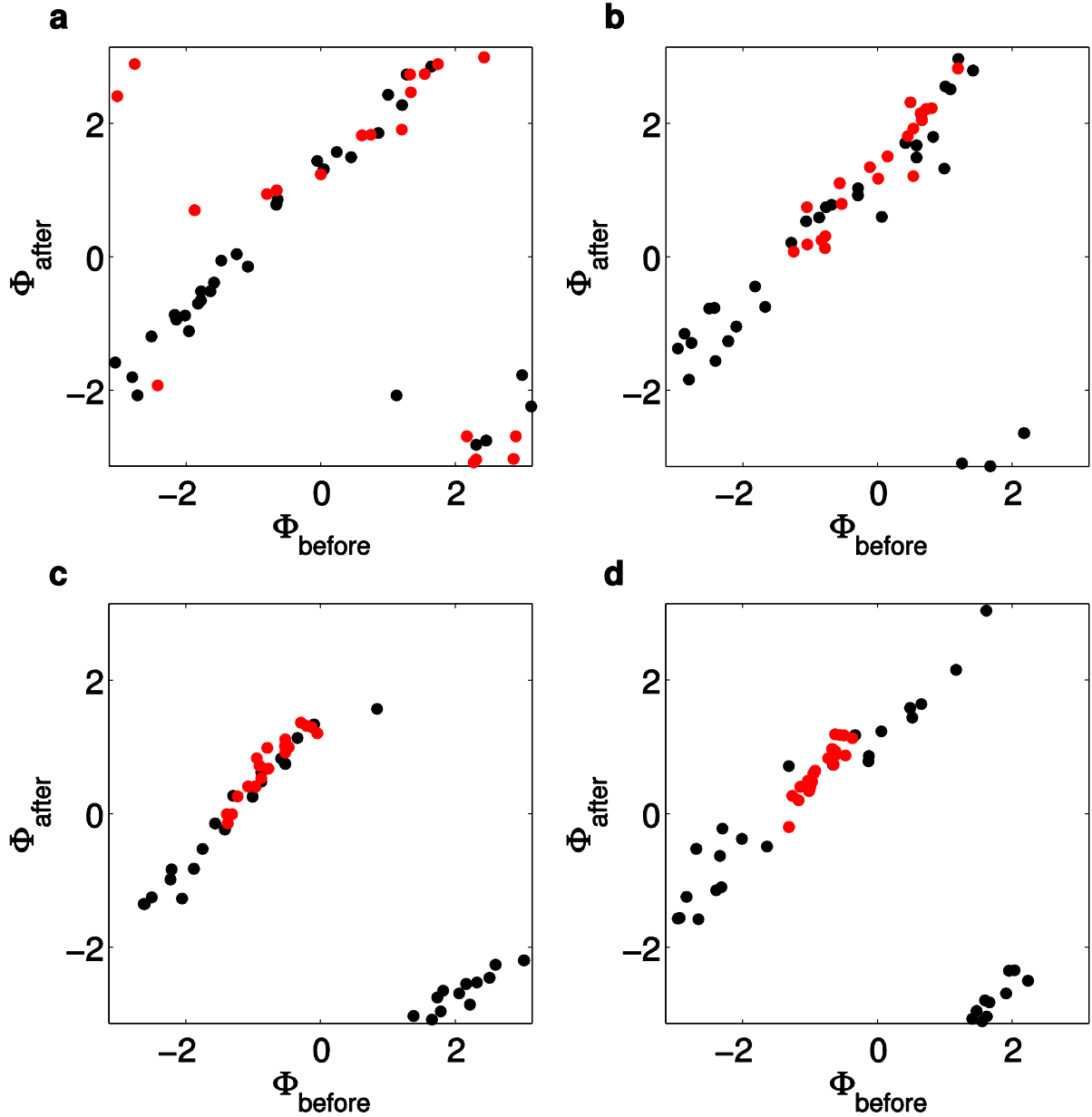

**Figure 3 Phases before and after perturbations.** Graphs showing the phases 9 s after cyanide removal/flow change as a function of the phases just before the removal/flow change for perturbations 1, 2, 4 and 8 in panel **(a)**-**(d)** respectively. Data from experiments where cells were entrained by cyanide removal is shown in red ($N$=20), while data from control experiments where only the flow rates were changed is shown in black ($N$=32). In panel **(a)** it can be seen that the phases before the first perturbation are evenly distributed for both data sets. In panels **(b)**-**(d)**, the phases of cells in the control experiments remain evenly distributed, while the distribution of phases of cells experiencing cyanide removal narrows to about 20% of the original distribution after perturbation 8. The graphs show that synchronising phase shifts are induced in the experiment where cyanide is removed and that most of the synchronising shifts occur at the first perturbations. For the control experiment, on the other hand, no phase shift dependence on the phase can be seen.

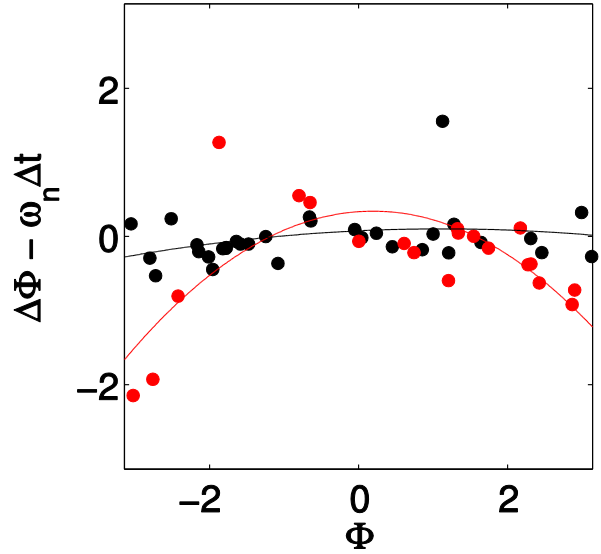

**Figure 4 Fits of corrected phase shifts.** Graph showing the phase shifts calculated from the phases shown in Figure 3a, corrected with the expected phase shift if no entrainment by the perturbations occur, $\omega_n \Delta t$, where $\omega_n$ is the frequency of the individual cells and $\Delta t$ is the time between the measurements of the phases before and after the perturbation. The two data sets are fitted to second-degree polynomials (see equation (1) and Table 1 for parameter values), which reveals a significant difference between the values of the quadratic terms. Cells that have phases close to -π or π become negatively phase shifted when cyanide is removed, while cells which have phases close to that of the perturbation remain relatively unaffected by the perturbation.

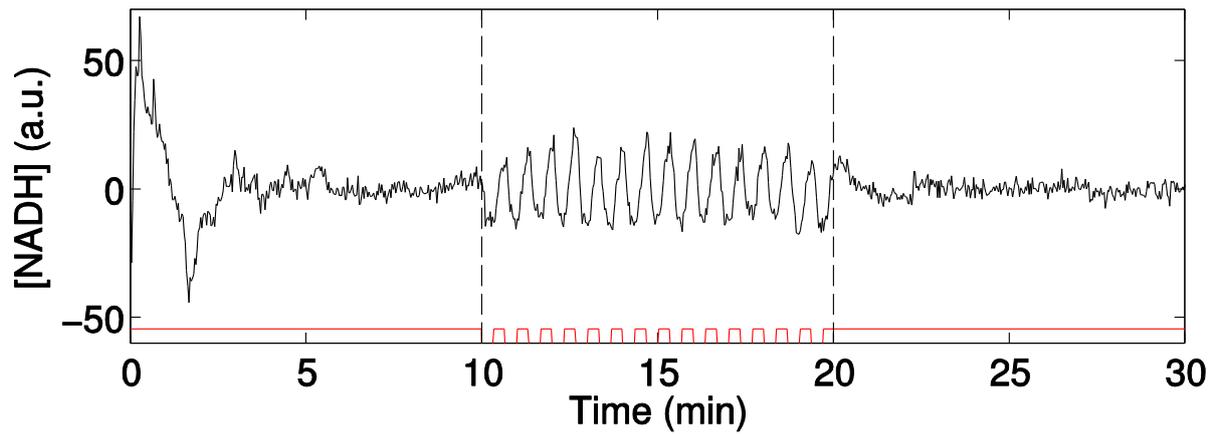

**Figure 5 Entrainment of non-oscillatory cells.** Graph showing the NADH fluorescence intensity of a single cell during the same experimental conditions as shown in Figure 1. Even though the cell is non-oscillatory for 20 mM Glc + 5 mM KCN, it still becomes entrained when cyanide is periodically removed (shown in red), and can thus contribute to a synchronised macroscopic response to perturbations. However, in contrast to the entrainment of oscillatory cells, which relies on phase synchronisation, entrainment of non-oscillatory cells requires a change also of the amplitude.

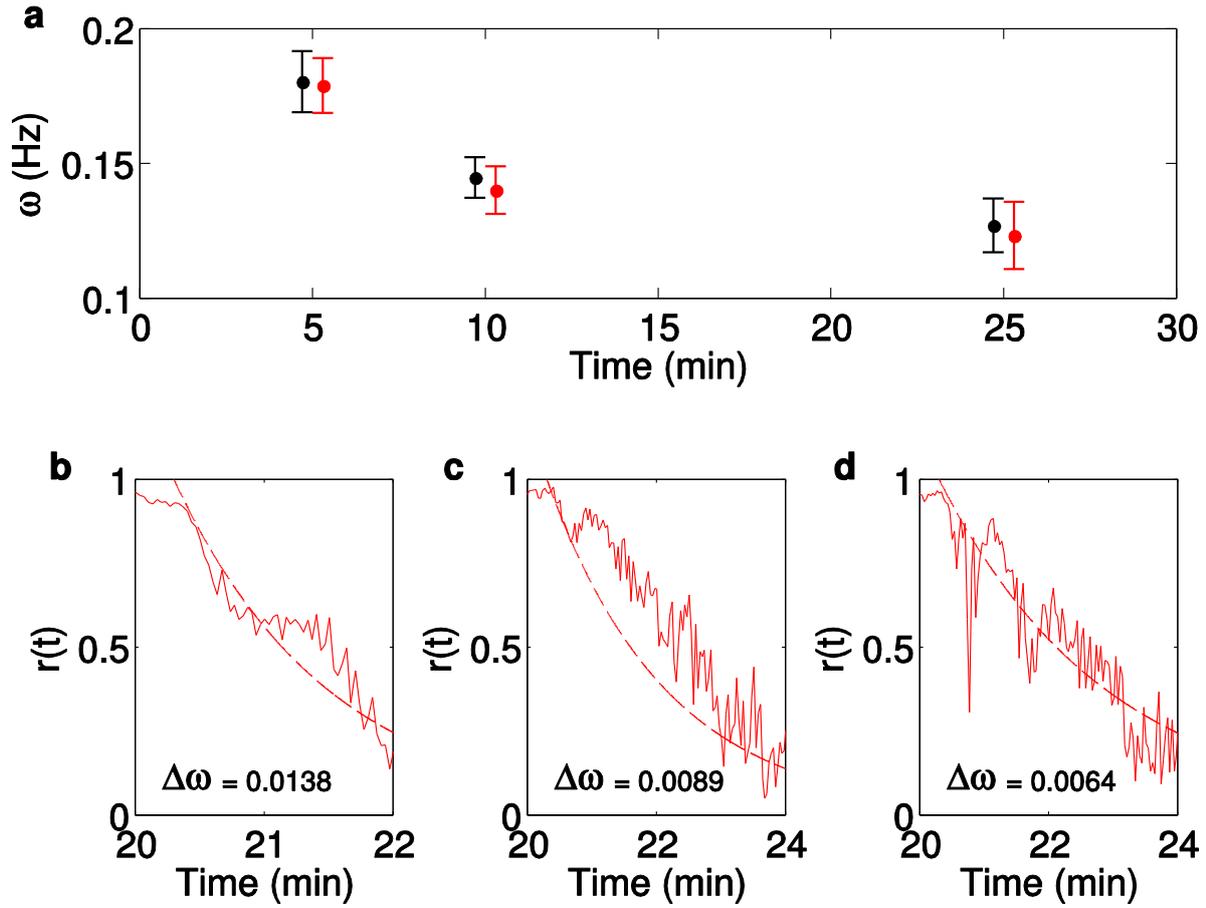

**Figure 6 Heterogeneity of the frequency of the oscillations and its influence on the decay of the order parameter.** (a) Graph showing the frequency of the oscillations expressed as mean ± standard deviation, where data from experiments where cells were entrained is shown in red ($N=24$), while data from control experiments is shown in black ($N=20$). No significant difference between the two data sets can be seen, neither with respect to the mean values of the frequency nor to their distribution. That entrainment occurs despite the large heterogeneity of phases indicate that the synchronisation mechanism is robust. In (b)-(d) the decay of the order parameter is shown for three different experiments where cells were entrained (solid lines, $N=14$, 10 and 10 respectively), together with exponential equations of the decay, where the rates are described by the distribution of frequencies $\Delta\omega$ of the oscillations in the respective experiments (dashed lines, see equation (3)). The decay was set to start 18 s after the end of the experimental perturbation at 20 min. It can be seen that the decay rate of the order parameter well can be estimated by the frequency distribution among the individual cells in the experiments.

**Table 1** Results of polynomial fit of the corrected phase shifts: $\Delta\phi - \omega_n \Delta t = c_2 \phi^2 + c_1 \phi + c_0$ at cyanide removal (20 cells) and for control experiments without cyanide removal (32 cells) see also Figure 4 and Methods. The constant term ($c_0$) and the linear slope ($c_1$) do not significantly differ from zero for either of the data sets. However, the most interesting parameter is the quadratic term ($c_2$), which does not significantly differ from zero for the control experiments, but has a significant non-zero value for the experiments where cells were entrained.

|  | Parameter estimates: mean value (95% confidence interval) | | |
| --- | --- | --- | --- |
|  | $c_2$ | $c_1$ | $c_0$ |
| Cyanide removal | -0.18 (-0.27, -0.09) | 0.07 (-0.07, 0.21) | 0.34 (-0.01, 0.78) |
| Control | -0.02 (-0.06, 0.02) | 0.04 (-0.02, 0.12) | 0.08 (-0.12, 0.27) |